\documentclass[aoas,MSNbibl,nameyear,dvips]{arximspdf}
\usepackage{dcolumn}
\usepackage{multirow}
\usepackage{graphicx}

\doi{10.1214/13-AOAS633} 
\volume{7}
\issue{3}
\pubyear{2013}
\firstpage{1684}
\lastpage{1708}

\makeatletter

\newcolumntype{d}[1]{D{.}{.}{#1}}

\renewcommand{\widehat}{\hat}

\newcommand{\bmu}{\mathbf{u}}
\newcommand{\bmv}{\mathbf{v}}
\newcommand{\T}{\mathbb{T}}

\makeatother

\begin{document}
\begin{frontmatter}

\title{Interpolation of nonstationary high frequency spatial--temporal
temperature data\thanksref{T1}}
\runtitle{Nonstationary spatial--temporal interpolation}

\thankstext{T1}{Supported by U.S. Department of Energy Grant DE-SC0002557.}

\begin{aug}
\author[A]{\fnms{Joseph} \snm{Guinness}\corref{}\ead[label=e1]{jsguinne@ncsu.edu}}
\and
\author[B]{\fnms{Michael L.} \snm{Stein}\ead[label=e2]{stein@galton.uchicago.edu}}
\runauthor{J. Guinness and M. L. Stein}
\affiliation{University of Chicago}
\address[A]{Department of Statistics \\
North Carolina State University \\
Raleigh, North Carolina 27695 \\
USA\\
\printead{e1}}
\address[B]{Department of Statistics \\
University of Chicago \\
Chicago, Illinois 60637 \\
USA\\
\printead{e2}} 
\end{aug}

\received{\smonth{9} \syear{2012}}
\revised{\smonth{11} \syear{2012}}

%
\begin{abstract}
The Atmospheric Radiation Measurement program is a U.S. Department of
Energy project that collects meteorological observations at several
locations around the world in order to study how weather processes
affect global climate change. As one of its initiatives, it operates a
set of fixed but irregularly-spaced monitoring facilities in the
Southern Great Plains region of the U.S. We describe methods for
interpolating temperature records from these fixed facilities to
locations at which no observations were made, which can be useful when
values are required on a spatial grid. We interpolate by conditionally
simulating from a fitted nonstationary Gaussian process model that
accounts for the time-varying statistical characteristics of the
temperatures, as well as the dependence on solar radiation. The model
is fit by maximizing an approximate likelihood, and the conditional
simulations result in well-calibrated confidence intervals for the
predicted temperatures. We also describe methods for handling
spatial--temporal jumps in the data to interpolate a slow-moving cold
front.\looseness=-1
\end{abstract}

%
\begin{keyword}
\kwd{Nonstationary process}
\kwd{spatial--temporal modeling}
\kwd{evolutionary spectrum}
\kwd{spatial--temporal jumps}
\end{keyword}

\end{frontmatter}

\section{Introduction}
When analyzing surface meteorological data, we often encounter
observations that are recorded at regular and frequent intervals in
time but at irregular and sparse locations in space. The data usually
come in the form of multiple time series, in which a small number of
lengthy time series are associated with fixed locations in space. The
methods developed in Stein (\citeyear{stein05,stein09}) exploit this structure in
constructing computationally efficient likelihood approximations for
data that may be modeled as a realization of a stationary
spatial--temporal Gaussian process or a very constrained type of
nonstationary Gaussian process. To model the temperature data that we
consider in this paper, a~more flexible nonstationary time series model
is needed to capture accurately the statistical properties of the
data,\vadjust{\goodbreak}
especially the dependence on solar radiation and time of day. The time
series model that we employ relies on the idea of the evolutionary
spectrum, as introduced by \citet{priestley65} and advanced
theoretically by Dahlhaus (\citeyear{dahlhaus96,dahlhaus12}). Through a simple
modification of the nonseparable spatial--temporal covariance function
in Stein (\citeyear{stein05,stein09}), we incorporate evolutionary spectra to
introduce nonstationary behavior in the time domain of the
spatial--temporal process. Using the computational techniques described
in \citet{guinness12}, we are able to efficiently compute and maximize
approximate Gaussian likelihoods for these data, which comprise 648,000
temperature observations. We also develop specialized methods for
modeling and interpolating spatial--temporal jumps in the data.

The goal of this work is to develop methods for producing accurate
space--time interpolations of temperature data at sites for which no
observations are available. This objective has a long history in the
meteorological literature, and many researchers have made important
contributions, including \citet{daly94}, who interpolate precipitation
data in the U.S., and \citet{hijmans05}, who interpolate several
climate variables at a high resolution with global land coverage. While
we have similar goals, our methods differ in that we characterize the
entire spatial--temporal statistical distribution of the temperature
process so that we may naturally incorporate uncertainty into the
interpolations. Specifically, we use the fitted statistical model to
produce a suite of simulated temperature data at unobserved locations
conditional on the observed temperature data, accounting for the
various uncertainties associated with the fitted model. Such
simulations are similar in spirit to the idea of multiple imputations
[\citet{rubinbook}] and have been referred to as data ensembles
[\citet{stein09,schneider06}]. In the geostatistics literature, this
approach is usually called geostatistical simulation
[\citet{geosimbook}, e.g.]. Because our statistical model tries
to reflect the nonstationary nature of the temperature records, the
resulting conditional simulations closely resemble the observed
temperature data; not only are the simulations accurate, but the
variance and the correlation structure are consistent with the
observations. Such simulations may be useful in any application that
requires a measure of uncertainty in addition to a point interpolation
of temperature. For example, a high resolution regional climate model
may require meteorological fields on a spatial grid as inputs, and the
simulations provide a way to propagate the uncertainty of the
interpolations through the climate model.

%
\begin{figure}[b]

\includegraphics{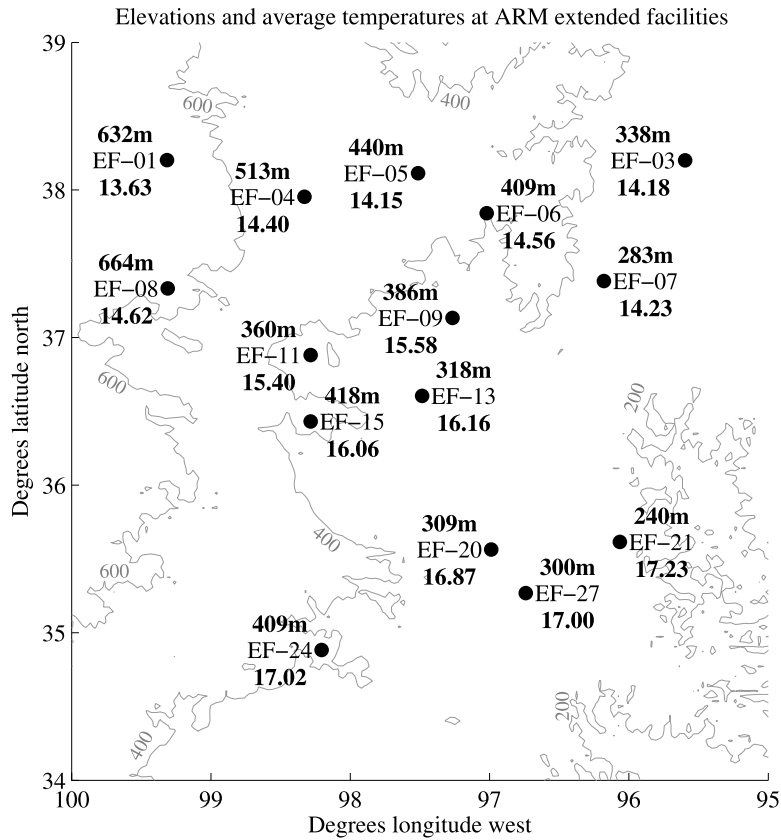}

\caption{Locations and elevations (meters) of the monitoring sites and
their October 2005 average temperatures ($^\circ\mbox{C}$). One
degree latitude
is approximately 111 km, and one degree longitude at 37 degrees latitude
is approximately 89 km. The central facility is located at site EF-13.}
\label{locaverages}
\end{figure}
%

\section{Atmospheric radiation measurement program data}

\begin{figure}

\includegraphics{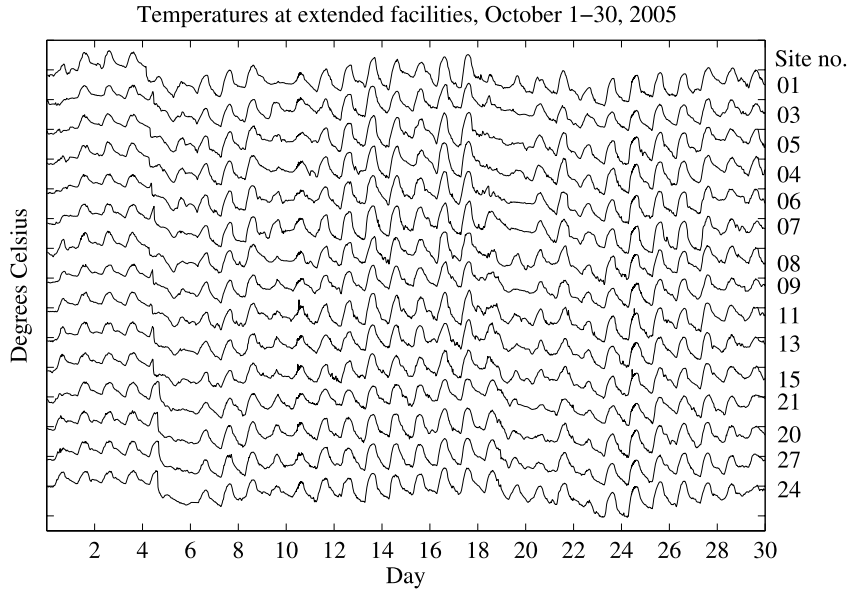}

\caption{Temperature record for October 1--30 at 15 ARM SGP extended
facilities. The tick marks on the vertical axis are separated by $20
^\circ\mbox{C}$, and the sites are ordered from north (top series) to south
(bottom series) and offset from each other by $20 ^\circ\mbox{C}$.}
\label{alltempsundiff}
\end{figure}

The data are provided by the Atmospheric Radiation Measurement (ARM)
Program, which was established by the U.S. Department of Energy in 1989
to study how weather processes, especially cloud formation, affect
global climate. The ARM program includes several mobile measurement
facilities, but the majority of the observations come from permanent
facilities in three primary regions: the North\vadjust{\goodbreak} Slope of Alaska, the
Tropical Western Pacific, and the Southern Great Plains (SGP). The SGP
field measurement site consists of a central facility in northern
Oklahoma and a modest number of extended facilities spread over 55,000
square miles in northern Oklahoma and southern Kansas. The data
analyzed in this paper were collected at the SGP site, and all of the
data may be accessed on the web at \url{http://www.archive.arm.gov}. The
extended facilities collect a host of meteorological observations,
including surface measurements of air temperature, air pressure,
relative humidity, and horizontal wind speed and direction. We analyze
the temperature data from 15 extended facilities from the first 30 days
of October, 2005, and we also make use of a solar radiation measurement
collected at the central facility. The positions and the elevations of
the 15 monitoring sites and the average temperatures from the first 30
days of October 2005 are plotted in Figure~\ref{locaverages}. The
temperature data are collected every minute to the nearest $0.01 ^\circ
\mbox{C}$.
Modeling and analysis of meteorological data at such a high temporal
frequency are of basic scientific interest, as evidenced by the
emergence of the field of micrometeorology, which aims to describe the
physical properties of atmospheric processes on very fine scales. In
addition, the high frequency interpolations may be of use as inputs to
high resolution regional climate models. When a regional climate model
has a spatial resolution of 10 km or less, it is typically run at time
steps of less than one minute. Such high resolution models include the
SPoRT-WRF model, which includes some 1 km resolution modeling
(\url{http://weather.msfc.nasa.gov/sport/modeling/}), and HIRLAM-B, which has
a target resolution of 2.5 km (\url{http://hirlam.org}). The ARM SGP data we
analyze are of high quality in that only 50 are missing out of 648,000,
and since the missing data are so few, we simply linearly interpolate
them in time. Figure \ref{alltempsundiff} plots the temperature time
series from each site.

\begin{figure}

\includegraphics{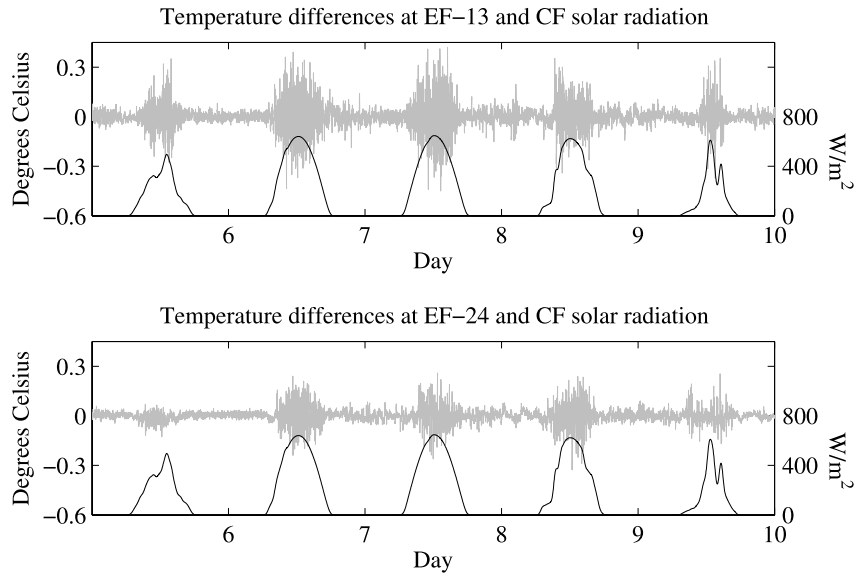}

\caption{Comparison between solar radiation measured at the central
facility and temperature differences measured at two sites, EF-13 and
EF-24. Site EF-13 is the site nearest the central facility, where the
solar radiation is measured.}
\label{diffrad}
\end{figure}

As expected, on most days the temperature data exhibit a clear diurnal
cycle in which the temperature begins to increase shortly after
sunrise, warms throughout the morning, and then begins to cool in the
late afternoon. In addition, the temperatures in the middle of the day
usually display more temporal variability than the nighttime
temperatures do. As a result, it is clear that a model for the
temperature (or differences of temperature) time series with a
stationary variance function is not sufficient for these data. Several
authors, including \citet{benth07} and \citet{campbell05}, have
addressed the issue of nonstationary variance in temperature
observations in the context of pricing financial products on the
weather derivatives markets. Their work mostly focuses on modeling
seasonal variances for mean daily temperatures, whereas here we
consider very high frequency data. For one-minute time resolution data,
there may be some hope in explaining the changing variability of the
temperature differences with other meteorological covariates. In fact,
incoming solar radiation, which is measured at the central facility at
one minute increments, has a strong connection to the temperature
variability, as seen in Figure \ref{diffrad}. We explore this
connection more closely in Section \ref{radsection}. In addition to
%
modeling a nonstationary variance, we model a nonstationary correlation
function, the details of which are discussed in Sections
\ref{lstatsection} and \ref{radsection}. Another conspicuous aspect of
the data is the existence of extreme jumps in temperature, most notably
during the fifth day, when the temperature at each site drops suddenly
over a short period of time. In Section \ref{jumpprocsection} we
develop methods for interpolating the jumps to unobserved sites.

\section{Statistical model}\label{statmodel}

We model the temperature process $X$ at time $t$ and location $\bmu$ as
%
\begin{equation}
X(t,\bmu) = m(t) + s(\bmu) + J(t,\bmu) + Y(t,\bmu),
\end{equation}
and we observe $X(t,\bmu_j)$ for $t=1,\ldots,T=43\mbox{,}200$ and for
$j=1,\ldots,n=15$, the number of observation locations. The function
$m$ is a nonrandom temporal mean function, $s$ is a spatial Gaussian
process, $J$ is a random spatial--temporal jump process, and $Y$ is a
nonstationary spatial--temporal Gaussian process. The temporal mean
function, $m$, is estimated by smoothing the time series of spatially
averaged temperatures. Specifically, we compute
%
\begin{equation}
\overline{X}(t) = \frac{1}{n} \sum_{j=1}^{n}
X(t,\bmu_j),
\end{equation}
and we smooth $\overline{X}(t)$ in time with a 20-minute-bandwidth
Daniell kernel applied three times to obtain $\hat{m}(t)$. The
bandwidth was chosen manually to balance smoothness and fit. We treat
$s(\bmu)$ as known at each observation location and set its value to
%
\begin{equation}
s(\bmu_j) = \frac{1}{T} \sum_{t=1}^T
X(t,\bmu_j) - \overline{\overline{X}},
\end{equation}
where $\overline{\overline{X}}$ is the grand mean of all observed
temperatures. The spatial Gaussian process model for $s$ allows us to
provide predictions and conditional simulations of the spatial means at
unobserved locations.

The random spatial--temporal jump process, $J$, is discussed in Section~\ref{jumpprocsection}.
The residual term, $Y$, is modeled as a
nonstationary spatial--temporal Gaussian process using techniques from
the spectral analysis of nonstationary time series. We discuss the
details of the time series model in Sections~\ref{lstatsection} and~\ref{radsection}, and we describe how we combine the time series
models into a spatial--temporal process model in Section \ref
{spatialmodelsection}.

\section{Spatial--temporal jump process}\label{jumpprocsection}
\begin{figure}

\includegraphics{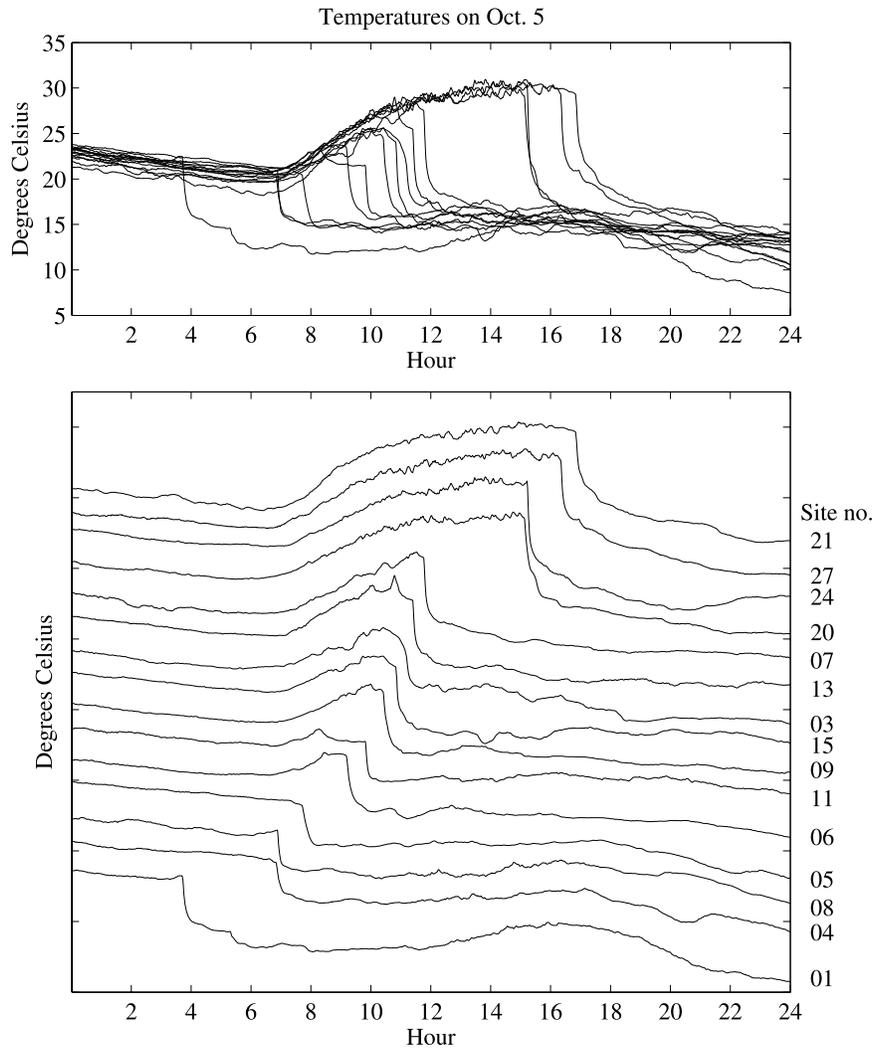}

\caption{Temperature record for October 5 at 15 ARM SGP extended
facilities. On the lower plot, the vertical axis tick marks are
separated by $10 ^\circ\mbox{C}$, and the temperatures are offset by
$4 ^\circ\mbox{C}$ per
site, ordered by the time of the extreme temperature drop.}
\label{day5undiff}
\end{figure}

It is clear that incorporating nonstationarity into the temporal domain
will provide a significant improvement to the fit. There is, however,
one important aspect of the data that remains difficult to capture if
we are to keep the amount of nonstationarity relatively constrained:
the extreme drop in temperature occurring on October 5, plotted in
Figure \ref{day5undiff}. On this day, each site experiences a rapid
drop in temperature of roughly 4 to $10 ^\circ\mbox{C}$ within a
period of no
more than 20 minutes, during which the first differences of temperature
can be as high as $2 ^\circ\mbox{C}$. In contrast, the differences
during the
rest of the month have a sample standard deviation of $0.06 ^\circ
\mbox{C}$. The
issue is further complicated by the fact that the temperature drops are
not simultaneous across the sites. Indeed, the site in the southeastern
corner of the region records its drop more than 12 hours after the site
in the northwestern corner records its drop. There seems to have been a
weather system moving slowly across the region on that day.

We do not think it would be a good idea to model the jumps as part of
the nonstationary spatial--temporal Gaussian process. For these data,
such a Gaussian process model would need to be sufficiently
pathological that it would violate the idea that the model is nearly
stationary on short time scales. A more suitable option is to fit a
random spatial--temporal jump process model in which the timing, sizes
and steepness of the jumps are random spatial processes. This not only
allows us to interpolate the jumps at unobserved sites but also
provides a way to model the uncertainty of the interpolated jumps.

\begin{figure}

\includegraphics{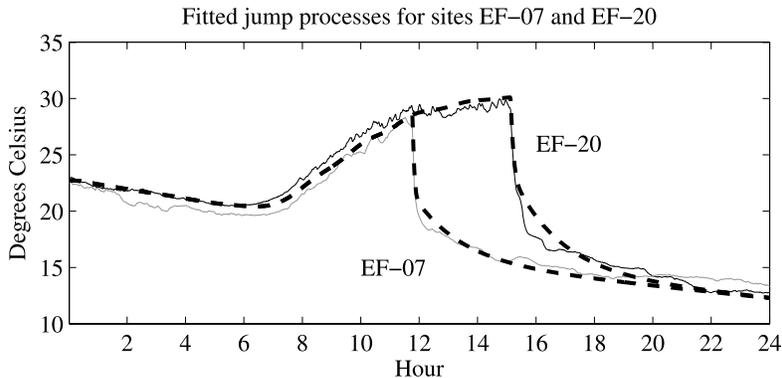}

\caption{Example of fitted jump processes at sites EF-07 and EF-20.
The fitted jump processes are the dashed lines, the temperatures from
site EF-07 are the solid gray line, and the temperatures from site
EF-20 are the solid black line.}
\label{fittedjump}
\end{figure}

Consider the following parametric form for the jump process:
%
\begin{equation}
J(t,\bmu) = \cases{ f\bigl(t; \tau(\bmu),D(\bmu), \lambda(\bmu)\bigr),
&\quad if
$1440\times4 < t \leq1440\times5$,
\cr
0, &\quad if otherwise,}
\end{equation}
so that the jump process is nonzero only on day 5, and where $f$ is a
nonrandom function of time given $\tau(\bmu)$, the time the jump
occurs, $D(\bmu)$, the size of the jump, and $\lambda(\bmu)$, the
steepness of the jump. We introduce randomness into the jump process by
modeling these parameters as random spatial processes, allowing us to
model the uncertainty in the timing, size and steepness of the jumps.
The idea is that if we can estimate these parameters using the data
from locations $(\bmu_1,\ldots,\bmu_n)$, we can conditionally
simulate the parameters at unobserved sites $(\bmu_{01},\ldots,\bmu
_{0m})$ via the spatial models and plug the conditionally simulated
parameters into $f$ to construct conditionally simulated jump processes
at the unobserved sites. Two examples of the fitted jump processes are
included in Figure \ref{fittedjump}.

Choosing the specific form for $f$ requires careful input from the
modeler, and we leave the details to Appendix \ref{appA}. The particular
formulation that we chose is highly specialized for these data, so we
do not propose that our analysis may be blindly applied to other
non-Gaussian features in the data or even other jumps. We do not see at
present a way to automate the modeling of this kind of feature in high
frequency meteorological data. However, we may in general be able to
apply the idea of fitting parametric functions to features in the data
at each site and then interpolating the feature to unobserved sites by
spatially interpolating the fitted parameters.

In addition to the extreme drops in temperature, sites EF-11 and EF-15
record short upward bursts in which the temperatures are raised by
several degrees Celsius and subsequently return to the pre-burst
temperature a short time later, usually within 30 minutes. The bursts
are recorded on two separate days, Oct. 11 and 25. Because the bursts
are short-lived and do not cause any major shift in the process, we
simply remove and replace them with linear functions connecting the
pre- and post-burst temperatures. We do not know what is causing the
bursts; however, it is interesting to note that the bursts occur at two
sites separated by 50 km and the burst at the northern site follows the
burst at the southern site by 75 minutes on day 11 and by 125 minutes
on day 25. In any case, since they are short-lived, and the
temperatures return to near the pre-burst temperatures, little is lost
in ignoring them. In contrast, the drop on Oct. 5 is sustained, so it
is not possible to ignore it.

Once the mean functions and the jump process have been estimated, we
may subtract them from the data and fit a nonstationary
spatial--temporal Gaussian process model to
%
\begin{equation}
\label{yhat} \widehat{Y}(t,\bmu_j) = X(t,\bmu_j) -
\hat{m}(t) - s(\bmu_j) - \widehat{J}(t,\bmu_j),
\end{equation}
where $\hat{m}(t)$ and $s(\bmu_j)$ are described in Section \ref
{statmodel}, and $\widehat{J}(t,\bmu_j)$ is the estimate of the jump
process, which is outlined in Appendix \ref{appA}. The details of the model for
$\widehat{Y}$ are to be described in the following sections.

\section{Locally stationary time series model}\label{lstatsection}
The spatial--temporal model for the residuals, which will be discussed
in Section \ref{spatialmodelsection}, specifies nonstationary marginal
time series models at fixed locations in space. Before describing the
spatial--temporal model, it is instructive to first introduce some
concepts from nonstationary time series. Dahlhaus (\citeyear{dahlhaus96,dahlhaus12})
introduced an asymptotic framework for nonstationary time series in
which the process may be considered stationary over short time scales,
thus allowing for consistent estimation of the model. Dahlhaus's work
builds on the idea of the evolutionary spectrum, introduced by \citet
{priestley65}. Specifically, let $T$ be a positive integer and $Z$ be a
complex orthogonal increment process on $\T$, the unit circle, subject
to the constraint that $Z(\omega) = Z(-\omega)^*$, where $^*$ denotes
the complex conjugate. Furthermore, let $A$ be a complex function on
$[0,1] \times\T$ with $A(u,\omega) = A(u,-\omega)^*$ for every $u$
and $\omega$, and $\int_{\T} |A(u,\omega)|^2 \,d\omega< \infty$ for
every $u$. Then
%
\begin{equation}
\label{lstatproc} X_T(t) = \int_{\T} A(t/T,
\omega) e^{i\omega t}\,dZ(\omega)
\end{equation}
is a real-valued nonstationary process on $t=1,\ldots,T$.

Estimation of the model amounts to estimation of $A$, which is a
time-varying transfer function, or, more frequently, of $|A|^2$, which
is usually called an evolutionary spectrum.
Dahlhaus (\citeyear{dahlhaus97,dahlhaus00}) proposes Gaussian likelihood
approximations that may be maximized over some parametric family for
$|A|^2$. \citet{guinness12} provide an improved Gaussian likelihood
approximation. Their likelihood approximates the integral in
(\ref{lstatproc}) with a sum over Fourier frequencies, $\omega_j =
2\pi
j/T$, so that the nonstationary time series vector $\mathbf{X}_T$ may be
written approximately as a linear transformation of a vector of
uncorrelated random variables~$\widehat{\mathbf{Z}}$,
%
\begin{equation}
\mathbf{X}_T \approx C_T(A)\widehat{\mathbf{Z}},
\end{equation}
where $C_T(A)$ is a $T\times T$ matrix with $(t,j+1)$th entry
$A(t/T,\omega_{j})\exp(i\omega_j t)$. Therefore, the linear
transformation $C_T(A)^{-1}\mathbf{X}_T$ is approximately decorrelating,
and the negative loglikelihood approximation has the form
%
\begin{equation}
l_T(A) = \frac{T}{2}\log(2\pi) + \log\bigl|\det C_T(A)
\bigr| + \frac
{1}{2}\bigl\|C_T(A)^{-1}\mathbf{X}_T
\bigr\|^2.
\end{equation}
The likelihood approximation effectively makes the same approximation
that the Whittle likelihood [\citet{whittle53}] makes for stationary
time series. In practice, computation of the log determinant part of
the likelihood is costly, so Guinness and Stein [(\citeyear
{guinness12}), Equation 10] provide
an approximation to that term. At present, their log determinant
approximation does not carry any known general theoretical guarantees,
but it works well and outperforms other existing approximations in many
examples. The quadratic form term may be computed efficiently using
iterative methods and a fast Fourier transform (FFT), especially when
$A$ takes the form
%
\begin{equation}
\label{lowrank} A(t/T,\omega) = \sum_{k=1}^K
m_k(t/T)\mu_k(\omega)
\end{equation}
with $K$ being a small integer [\citet{guinness12}, Section 2]. We will
always make the additional assumption that $m_k$ and $\mu_k$ are
nonnegative. The form in (\ref{lowrank}) includes the stationary model
as a special case with $K=1$ and $m_1$ constant. It also includes the
uniformly modulated model as described by \citet{priestley65}, with
$K=1$ and $m_1$ not constant. When $K>1$, we refer to the $\mu_k$'s as
``regimes,'' so that at any time $t/T$, the transfer function is a
positive linear combination of the various regimes. This representation
is similar to the approach taken in hidden state modeling, where the
process can be described at any time by a particular state or a
superposition of several states. \citet{fuentes02} proposed a similar
formulation in which the nonstationary process is a spatially-varying
superposition of several independent stationary processes.

For a spectral analysis of time series, it is often advantageous to
prewhiten the data to smooth out peaks in the power spectrum [\citet
{priestleybook}]. The spectra of the high frequency temperature time
series tend to have peaks at the origin, and one way to smooth out a
peak at the origin is to prewhiten the data with the difference filter,
that is, perform a spectral analysis on the first differences, $\Delta
X(t) = X(t) - X(t-1)$. Differencing is exactly a transformation to
white noise when the time series is Brownian motion. For the
temperature time series, we argue that differencing is too strong a
filter because temperatures do exhibit some mean-reversion over monthly
time scales, whereas Brownian motion does not over any time scale.
Therefore, we propose to perform the analysis on partial differences,
defined as\looseness=-1
%
\begin{equation}
\Delta_\alpha X(t) = X(t) - \alpha X(t-1).
\end{equation}\looseness=0
The parameter, $\alpha\in[0,1]$, which controls the amount of
differencing, may be estimated with maximum likelihood. Partial
differencing also stabilizes the simulated time series upon the
operation of undifferencing, which has a huge impact on the usefulness
of the conditional simulations (see Appendix~\ref{appC}).

\section{Solar radiation and choosing regimes}\label{radsection}
One striking feature of the data set is the relationship between the
temporal variability in the first differences of temperature and the
amount of incoming solar radiation: large amounts of sunshine usually
result in more variable temperatures. Figure \ref{diffrad} shows an
example of the relationship. Meteorologists have hypothesized that
convective forces in the atmosphere, which are most active during the
middle of the day, may be driving the variability. One theory predicts
that the standard deviation of surface air temperature changes should
be linearly related to the heat flux at the earth-atmosphere boundary
raised to the $2/3$ power [\citet{aryabook}, page 183]. We have found
that, for these data, the standard deviation of the partial differences
is approximately linearly related to incoming solar radiation. We use
the radiation as a covariate to attempt to standardize the temporal
variance in the differences. In order to quantify the ability of solar
radiation to explain the time-varying variance, we fit three
spatial--temporal models to the first differences of temperature at the
15 monitoring locations: (a) a stationary model, as described in
\citet{stein09}, (b) a model that is stationary except for a separate
variance for day and night, and (c) a model that is stationary except
for a standard deviation that depends linearly on smoothed solar
radiation measured at the central facility (the smoothing is described
below). The day/night model increased the loglikelihood by 67,543 units
over the stationary model, and the radiation-dependent model increased
the loglikelihood by 104,286 units over the stationary model, so by any
measure such as AIC or BIC, the additional variance parameters in both
models are highly meaningful, and solar radiation explains the variance
significantly better than does time of day alone.

To get a preliminary estimate for the relationship between the
differences and solar radiation, we find $\hat{a}_0$ and $\hat{a}_1$
by fitting the model
%
\begin{equation}
\label{prelimrad} \Delta_\alpha\widehat{Y}(t,\bmu) =
\bigl[a_0 + a_1 r\bigl(t-\theta\mathbf{u}^\prime
\bolds{\phi}\bigr)\bigr]\varepsilon(t,\bmu),
\end{equation}
where $\varepsilon(t,\bmu)$ are i.i.d. $N(0,1)$. The smoothed
radiation, $r(t)$, is obtained by applying a left-sided,
10-minute-bandwidth Daniell kernel three times to the observed
radiation. The parameters $\theta$ and $\bolds{\phi}$ control the size
and direction of the phase. For radiation from the sun, the phase
should move west [$\bolds{\phi} = (-1,0)'$] at a speed of $\theta=4$
minutes per degree longitude. Later, we will refine the estimates of
$a_0$ and $a_1$ by maximizing the full approximate likelihood over
them. Figure \ref{diffrad} shows first differences of temperature at
site EF-13, which is located at the central facility where the
radiation is measured, and at site EF-24, which is more than 200 km away
from the central facility. It is not surprising that the relationship
between variability and solar radiation is weaker when the site is far
away from where the radiation measurement was taken. If one-minute
resolution radiation data were available at every site, using that data
would be preferable, but the extended facilities record radiation data
only on the hourly resolution. There is also a matter of convenience:
if central radiation measurements are used, there is no need to
interpolate the radiation to unobserved sites. Despite these
shortcomings, since a single covariate explains so much of the changing
variability, it is worthwhile to include it in the model.

\begin{figure}

\includegraphics{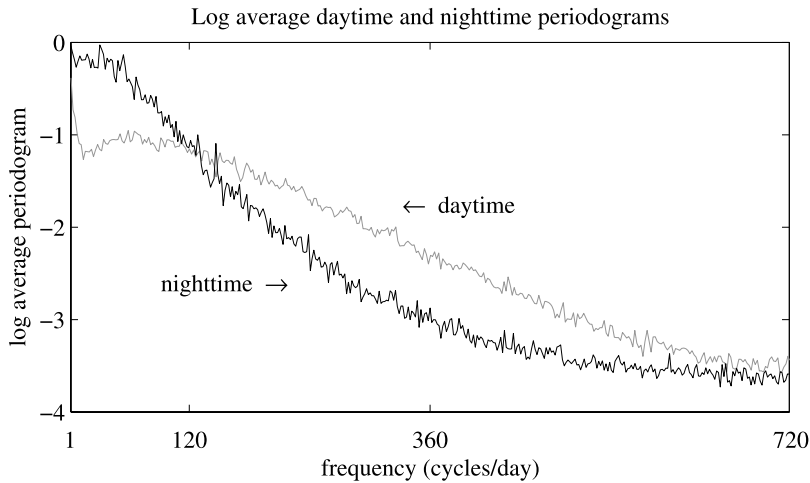}

\caption{Log average daytime (gray) and nighttime (black) periodograms
computed with no sunrise or sunset offsets. Daytime (nightime) average
is taken over all daytime (nighttime) blocks over all sites.}
\label{avgpgrams}
\end{figure}

Another feature of the data, although less striking than the radiation
phenomenon, is that the spectral properties of the time series differ
depending on the time of day. Furthermore, the difference cannot be
explained with a uniform modulating function; the mean-zero daytime
process is not simply\vspace*{1pt} a multiple of the mean-zero nighttime process. To
illustrate this point, after dividing $\Delta_1 \widehat{Y}(t,\bmu)$
by $\hat{a}_0 + \hat{a}_1 r(t-\theta\bmu^\prime\bolds{\phi})$, we
partition the data into daytime and nighttime blocks according to
approximate local sunrise and sunset times,\setcounter{footnote}{1}\footnote{The exact time of
each day's local sunrise and sunset at the central facility was
recorded, and then the approximate local sunrises and sunsets at the
other sites were assumed to be offset by four minutes per degree
longitude from the sunrise and sunset at the central facility on each
day. At the beginning of the month, this approximation is nearly exact,
as Oct. 1 is close to the equinox. At the end of the month, the
approximation is off by roughly four minutes for the most northern
site, which for our purposes is an acceptable error.} compute
periodograms for each block of data, and average all the periodograms
from all the daytime blocks from all the sites to get an average
daytime periodogram, and we average all the periodograms from all the
nighttime blocks from all the sites to get an average nighttime
periodogram. We plot the two averages in Figure \ref{avgpgrams}. It is
clear that the daytime spectrum cannot be modeled as a multiple of the
nighttime spectrum. This means that not only does the variance of the
process change over time, but the correlation structure changes as well.

Based on this evidence, we propose that the time series model requires
at least two regimes [$K=2$ in (\ref{lowrank})], and a simple
candidate model is one with a ``daytime'' regime and a ``nighttime''
regime. It is not clear how one would automatically estimate the two
regimes, $\mu_1$ and $\mu_2$, via maximum likelihood. However, if we
take the regimes to be fixed and known, it is possible to maximize the
likelihood over a family of modulating functions. Therefore, we
estimate the regimes beforehand and treat them as fixed when we
maximize the likelihood. In short, after defining the time at which the
day starts and ends, which we do not assume is exactly at local sunrise
and sunset, we estimate the regimes by fitting B-splines to the log of
the average daytime and nighttime periodograms. One issue with the
daytime and nighttime periodograms is that they do not provide
information about the very low frequency behavior of the process
because the periodograms are taken over short blocks of data. To
estimate the very low frequency behavior, we borrow some information
from the periodograms of the entire month of data. The details of the
estimation of the regimes are left to Appendix~\ref{appB}.\looseness=1

With the regimes fixed, we introduce a parametric form for the
modulating functions, $m_1$ and $m_2$, in (\ref{lowrank}). We
partition the time interval into $B$ blocks, and we assume that, aside
from a radiation-dependent uniform modulation, the evolutionary
spectrum is constant in time on the blocks
%
\begin{eqnarray}
\label{monthlyA} M(t/T,\omega) &=& \sum_{k=1}^2
\sum_{b=1}^{B} w_{kb}
\mathbf{1}_b(t/T) \mu_k(\omega),
\nonumber\\[-8pt]\\[-8pt]
A(t/T,\omega) &=& \bigl(a_0 + a_1 r(t)\bigr)M(t/T,
\omega),
\nonumber
\end{eqnarray}
where $w_{kb}>0$, the $\mu_k$'s are the regimes, and $\mathbf{1}_b(x)$ is
an indicator function that equals 1 if $x \in$ block $b$ and 0
otherwise. Therefore, the parameters $w_{kb}$ determine the weight
assigned to regime $k$ in block $b$. The model in (\ref{monthlyA})
bears some resemblance to a cyclostationary model, in which the process
is nonstationary within each day, but the nonstationary structure
exactly repeats itself every day [\citet{gardner06}]. Here, we do not
require the spectrum to be cyclic, but we do enforce a relatively
simple dependence on just two ``regimes.''

Because the regimes are taken to be fixed, a maximum approximate
likelihood procedure involves choosing the linear radiation
coefficients, the weights assigned to the regimes within each block,
and the positions of the changepoints defining the blocks (in addition
to fitting the spatial dependence structure, which is discussed in
Section \ref{spatialmodelsection}). \citet{guinness12} implemented a
genetic algorithm to find the changepoints, which works well when there
are 1--10 changepoints. Here, however, where there are 60 changepoints
per site, a more constrained search is necessary. We propose that it is
reasonable to impose that the changepoints occur a fixed amount of time
before or after local sunrise and sunset at each site and on each day.
This constraint leaves us with just two changepoint parameters to
estimate, so we may perform a grid search to find the best sunrise and
sunset offsets according to the associated likelihoods. For each choice
of sunrise and sunset offsets, we re-estimate the daytime and nighttime
regimes as outlined in this section and maximize the likelihood over
the radiation coefficients, the $w_{kb}$'s, and the parameters
describing the spatial dependence to be described in Section \ref
{spatialmodelsection}.

\section{Nonstationary spatial--temporal Gaussian process model}\label
{spatialmodelsection}
Recent work on incorporating nonstationarity into Gaussian process
models has come in several forms. One of the approaches is to assume
that data come from an isotropic random field on an unobserved domain
$E$, and the domain on which we observed the data, $D$, is related to
$E$ via an unknown invertible mapping. Early work on this approach is
due to \citet{sampson92}, and more recent advances include
\citet{anderes08} and \citet{anderes09}. Several authors have
constructed nonstationary processes via the convolution of a family of
independent (of each other) stationary Gaussian processes.
Nonstationarity enters when the convolution kernels are allowed to vary
across the domain. \citet{higdon99} use squared exponential covariance
functions for the component processes, and \citet{paciorek06} provide
an approach to include the Mat\'{e}rn class of covariance functions.
\citet{fuentes02} expresses the nonstationary process at each spatial
location as a spatially-varying linear combination of stationary
processes and estimates the model in the spectral domain. Others have
modeled the nonstationary process in the spectral domain through the
use of spatially- or spatial-temporally-varying spectral densities,
analogous to the evolutionary spectrum approach studied in Section
\ref{lstatsection}. \citet{fuentes07} propose a class of nonseparable
and nonstationary covariance functions through the use of a spectrum
that varies across space and time.

The stationary, nonseparable covariance function in \citet{stein09} is
specially suited to facilitate fast approximate likelihood computations
when the data are collected at a sparse set of fixed monitoring
stations at regular intervals in time. In particular, he considered
stationary spatial--temporal covariance functions of the form
%
\begin{equation}
\label{statspacetime} K(t,\mathbf{u}) = \int_{-\pi}^{\pi}
S(\omega) R\bigl(|\bmu|/\gamma(\omega)\bigr)e^{i \omega(t-\theta\bmu'
\bolds{\phi})} \,d\omega,
\end{equation}
where $S$ is a spectral density, $R$ is a one-dimensional correlation
function invoking spatial coherence, $\gamma$ is a positive function
allowing the coherence to vary with temporal frequency, $\theta$ is a
scalar, and $\bolds{\phi}$ is a $2\times1$ vector. The parameters
$\theta$ and $\bolds{\phi}$ give the size and direction of the phase shift.

We construct a process that can be considered nonstationary in time by
allowing the spectrum in (\ref{statspacetime}) to depend on time.
Rather than writing the covariance function, we give a spectral
representation of the process,
\[
\Delta_\alpha Y(t,\bmu) = \int_{-\pi}^\pi A
\bigl(\bigl(t-\theta\bmu' \bolds{\phi}\bigr)/T,\omega
\bigr)e^{i\omega t}\,dZ_{\bmu}(\omega).
\]
Because the temporal argument of $A$ is shifted based on the location
$\bmu$, we make the slight modification that $A$ is a function defined
on $\mathbb{R} \times\T$. In practice, the phase shift parameters
have a very small effect on the likelihood, but rather than setting
$\theta=0$ and ignoring the phase, the default phase should correspond
to a shift to local time because we know the temperature process is
dependent on the time of day, so we fix $\bolds{\phi} = (-1,0)'$, and
$\theta=4$ minutes per degree longitude. The orthogonal increment
process, $Z_{\bmu}$, has spatial coherence
\[
E\bigl(dZ_{\bmu}(\omega)\,dZ_{\bmv}(\omega)^*\bigr) = R \biggl(
\frac{|\bmu-
\bmv|}{\gamma(\omega)} \biggr)e^{-i\theta\omega(\bmu- \bmv)'\bolds
{\phi}}\,d\omega,
\]
where $R$ is again a one-dimensional correlation function. We take
$R(d) = \exp(-|d|)$, and use the convention that $R(0/0)=1$. We
parameterize $\gamma(\omega)$ as a B-spline whose derivative at
$\omega=0$ is 0, equals 0 for all $|\omega|>\omega_0$, and whose first
and second derivatives at $\omega=\omega_0$ are 0. We choose $\omega_0$
to be 48 cycles per day, and we place the interior knots in the
B-spline at (1, 3, 6, 12, 24) cycles per day, leaving us with 5 basis
functions in the B-spline representation for $\gamma$. To ensure
$\gamma(\omega) \geq0$, we require that the B-spline coefficients be
greater than 0. The purpose of $\gamma$ is to allow the low (temporal)
frequency fluctuations of the process to be more strongly correlated
across space than the high frequency fluctuations of the process are,
a~phenomenon observed in the analysis of the Irish wind data
[\citet{stein05,haslett89}] and the ARM SGP air pressure data
[\citet{stein09}]. By setting $\gamma(\omega) = 0$ for all
$|\omega| > \omega_0$, we are assuming that there is no spatial
coherence above the 48 cycles per day frequency. This assumption is not
unrealistic, at least at the available spacings between monitoring
sites for these data, and it provides significant computational
advantages in the likelihood approximation. \citet{stein09} also
considered a spatial nugget effect, but for these data and choice of
coherence function, a nugget did not improve the fit, so we ignore it
here.

In principle, one could allow the spectrum to depend on space as well.
In fact, our spectrum does depend on space through the phase shift, but
this is a very constrained dependence. For the ARM temperature data,
which have just 15 spatial locations, we do not attempt to model any
nonstationary aspects of the spatial covariance function. However, in
the temporal domain, where we have 43,200 observations over the month
per site, introducing nonstationarity has a substantial impact on the
model fit.

\section{Likelihood approximation and model fitting}\label{liksection}

We fit a spatial--temporal model to the data with two sites removed,
site EF-08, which is on the western edge of the observation domain, and
EF-09, which is near the center of the observation domain. We refer to
their locations as $\bmu_{01}$ and $\bmu_{02}$. Given fixed values of
the sunrise and sunset offsets and the differencing parameter, our
method for fitting the spatial--temporal model proceeds as follows:
\begin{enumerate}
\item Estimate the temporal and spatial mean functions (Section
\ref{statmodel}) and the jump process parameters (Section \ref
{jumpprocsection}).

\item Construct $\widehat{Y}(t,\bmu_j)$ as in equation (\ref{yhat}).

\item Preliminarily estimate $a_0$ and $a_1$ by fitting the model in
(\ref{prelimrad}) to obtain $\tilde{a}_0$ and~$\tilde{a}_1$.

\item Divide each time series $\Delta_\alpha\widehat{Y}(t,\bmu_j)$
by $\tilde{a}_0 + \tilde{a}_1 r(t)$ and estimate the regimes using
the average daytime and nighttime periodograms, where the start of the
daytime and nighttime blocks are determined by local sunrise and sunset
plus the offsets.

\item Maximize the approximate likelihood, which will be described at
the end of this section, simultaneously over $a_0$, $a_1$, the 122
$w_{bk}$ parameters and the 5 parameters describing the B-spline
representation for $\gamma(\omega)$.
\end{enumerate}

\begin{figure}

\includegraphics{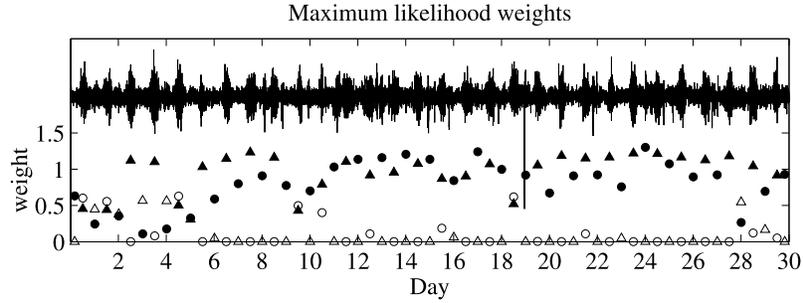}

\caption{Maximum likelihood estimates of the weights assigned to the
daytime (triangles) and nighttime (circles) regimes within each block.
The filled markers indicate regimes that match the time of day, that
is, if it is daytime, then the triangles are filled, and if it is
nighttime, the circles are filled, so we expect the filled markers to
receive more weight (e.g., we expect the daytime blocks to receive more
weight from the daytime regime). We also plot the time series of
partial differences for one of the sites, EF-27.}
\label{weights}
\end{figure}

\begin{figure}[b]

\includegraphics{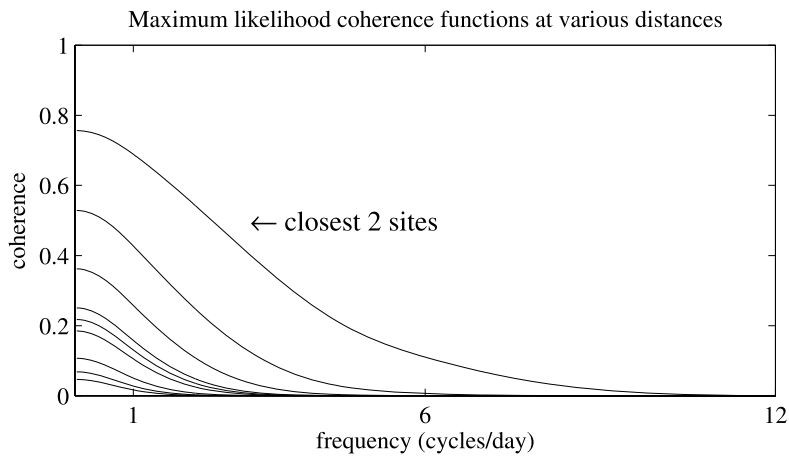}

\caption{Maximum likelihood estimate of the coherence as a function of
frequency. Each line corresponds to the coherence at a fixed spatial
distance. The coherence function was allowed to be nonzero at up to 48
cycles per day.}
\label{coherence}
\end{figure}

To speed the optimization, we compute analytic gradients of the
likelihood with respect to $a_0$, $a_1$ and each $w_{bk}$, and we
compute finite difference gradients with respect to the
$\gamma(\omega)$ parameters. It should also be noted that the model in
(\ref{monthlyA}) is overspecified in that we may multiply the radiation
coefficients by a constant and divide all the weights by the same
constant and arrive at the same function $A$. In the optimization
procedure, we fit the overspecified model,\vadjust{\goodbreak} as the naturally specified
model tends to get stuck in local minima. For fixed values of the
offsets and differencing parameter, the time it takes to complete the
optimization will vary but usually takes on the order of 2--3 hours on a
single processor. In Figure \ref{weights} we plot the maximum
likelihood weights, $w_{bk}$. As expected, the daytime blocks usually
receive more weight from the ``daytime'' regime, and the nighttime
blocks usually receive more weight from the ``nighttime'' regime,
although there are some exceptions. The maximum likelihood coherence,
$C(|d|/\gamma(\omega))$, as a function of frequency at various
distances is plotted in Figure \ref{coherence}.\vadjust{\goodbreak} The fitted model for
the temperature data exhibits much weaker spatial coherence than is
estimated for the pressure data in \citet{stein09}.

To explore the effect of the sunrise and sunset offsets, we repeat the
optimization first over a coarse grid and then over a finer grid of
sunrise and sunset offsets (with $\alpha$ fixed at 0.99). Once we have
obtained results on the finer grid, which has 10 minute spacings, we
fit a quadratic to the points near the maximum and then repeat the
optimization procedure at several points near the maximum of the
quadratic fit. The optimization over a grid is easily parallelizable
because we can assign each grid point to a processor, so the grid
search is relatively fast when one has access to multiple processors.
The results of the grid search are given in Table \ref{gridsearch}.
The offset parameters have a big effect on the likelihood, and we find
that sunrise offset $111$ minutes and sunset offset $-125$ minutes give
the highest maximum approximate likelihood; the temperature process
seems to undergo a change roughly two hours after sunrise and two hours
before sunset.

\begin{table}
\caption{Results of the grid search over sunrise and sunset offsets
(minutes) with differencing parameter $\alpha=0.99$. Table entries
refer to difference in loglikelihood from that found at sunrise offset
$111$ and sunset offset $-125$. The units are thousands of
loglikelihood units}\label{gridsearch}
\begin{tabular*}{\tablewidth}{@{\extracolsep{\fill}}l c c c c c c@{}}
\hline
\multicolumn{1}{@{}l}{\multirow{1}{32pt}[-7.8pt]{\textbf{Sunrise offset}}}
&\multicolumn{6}{c@{}}{\textbf{Sunset offset}}\\[-4pt]
&\multicolumn{6}{c@{}}{\hrulefill}\\
&\multicolumn{1}{c}{$\bolds{-200}$}&\multicolumn{1}{c}{$\bolds{-150}$}
&\multicolumn{1}{c}{$\bolds{-100}$}&\multicolumn{1}{c}{$\bolds{-50}$}
&\multicolumn{1}{c}{\textbf{0}}&\multicolumn{1}{c@{}}{\textbf{50}}\\
\hline
$-$50 & 10.05 & 8.32 & 8.03 & 8.30 & 8.22 &
10.34\\
\hphantom{0$-$}0 & \hphantom{0}8.43 & 6.47 & 6.31 & 6.92 & 7.03 & \hphantom{0}9.55\\
\hphantom{$-$}50 & \hphantom{0}5.96 & 3.51 & 3.41 & 5.00 & 5.53 & \hphantom{0}8.38\\
\hspace*{2.2pt}100 & \hphantom{0}3.78 & 0.71 & 0.51 & 3.43 & 5.03 & \hphantom{0}7.80\\
\hspace*{2.2pt}150 & \hphantom{0}4.28 & 1.45 & 1.68 & 4.67 & 6.31 & \hphantom{0}9.19\\
\hspace*{2.2pt}200 & \hphantom{0}7.22 & 5.39 & 6.07 & 7.73 & 8.63 & 11.81
\\[-4pt]
&\multicolumn{6}{c@{}}{\hrulefill}\\
&\multicolumn{1}{c}{$\bolds{-150}$}&
\multicolumn{1}{c}{$\bolds{-140}$}&\multicolumn{1}{c}{$\bolds{-130}$}
&\multicolumn{1}{c}{$\bolds{-120}$}&\multicolumn{1}{c}{$\bolds{-110}$}
&\multicolumn{1}{c@{}}{$\bolds{-100}$}
\\[-4pt]
&\multicolumn{6}{c@{}}{\hrulefill}\\
\hphantom{$-$}80 & 1.426 & 1.047 & 0.940 & 0.948 & 1.062 &
1.243\\
\hphantom{$-$}90 & 1.050 & 0.648 & 0.533 & 0.537 & 0.655 &
0.848 \\
\hspace*{2.2pt}100 & 0.712 & 0.302 & 0.177 & 0.176 & 0.303 &
0.506 \\
\hspace*{2.2pt}110 & 0.603 & 0.187 & 0.069 & 0.068 & 0.205 &
0.425 \\
\hspace*{2.2pt}120 & 0.682 & 0.278 & 0.162 & 0.174 & 0.326 &
0.566 \\
\hspace*{2.2pt}130 & 0.737 & 0.345 & 0.241 & 0.264 & 0.438 &
0.694\\
\hline
\end{tabular*}
\end{table}

\begin{figure}

\includegraphics{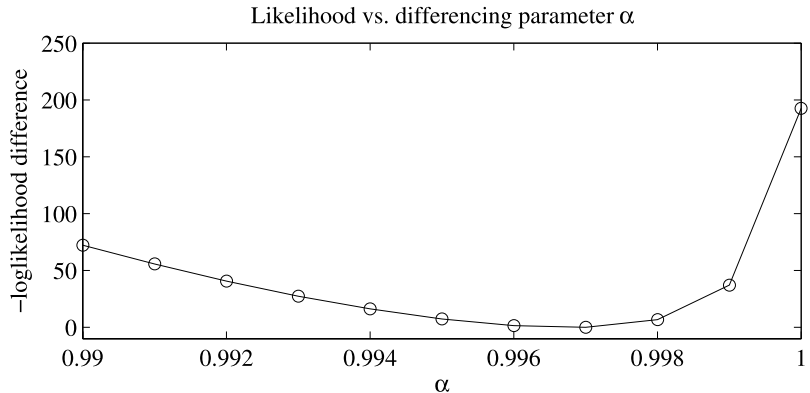}

\caption{Results of a grid search over $\alpha$, the differencing
parameter. Here we plot the difference from the loglikelihood maximized
when $\alpha=0.997$. We use sunrise offset 111 and sunset offset $-$125.}
\label{alphalik}
\end{figure}

One may also attempt to obtain a higher likelihood by altering $\alpha
$, the amount of partial differencing. Assuming that the differencing
does not alter the locations of the changepoints of the process, we may
use the estimates of the sunrise and sunset offsets obtained
previously, then repeat the likelihood optimization as before over a
grid of differencing parameters. In Figure \ref{alphalik} we plot the
maximum approximate likelihood as a function of $\alpha$ for sunrise
offset $111$ and sunset offset $-125$. Choosing $\alpha=0.997$ gives
the highest maximum approximate likelihood.

The Gaussian likelihood approximation for the spatial--temporal data
proceeds in much the same way as it does for the time series data.
After the mean functions and the estimated jump process have been
removed from the data, we construct $\Delta_\alpha\widehat{\mathbf{Y}}_j$,
which is the vector of partial differences of the residuals at site
$\bmu_j$, for $j=1,\ldots,n-2=13$, the number of observed sites. For
some choice of $A$, and for each site $\bmu_j$, we compute the
approximate transformation to independence $\widehat{\mathbf{Z}}_{j} =
C_T(A)_{j}^{-1}\Delta_\alpha\widehat{\mathbf{Y}}_j$, where the subscript
$j$ on $C_T(A)_{j}$ reminds us that the phase shift in the temporal
argument of $A$ depends on location $\bmu_j$ (we omit from the notation
the dependence on $\theta$ and $\bolds{\phi}$, which are taken to be
fixed). The inverse transformation is efficiently computed using an
iterative algorithm and the FFT, as described in \citet{guinness12}.
Then the approximate covariance matrix of
$(\widehat{\mathbf{Z}}_{1},\ldots,\widehat{\mathbf{Z}}_{n})$ is block diagonal
with block sizes equal to the number of spatial locations because we
allowed for spatial correlation in the orthogonal increment process.
Inverting this block diagonal covariance matrix is much easier than
inverting the full covariance matrix of the differences of the
residuals, especially since we assume that the spatial correlation in
$Z_{\bmu}$ is zero for all frequencies greater than 48 cycles per day.
The log determinant term is computed as in \citet{guinness12} with the
modification that the spatial correlation of
$(\widehat{\mathbf{Z}}_{1},\ldots,\widehat{\mathbf{Z}}_{n})$ must be accounted
for, but this is also quite simple because the log determinant of a
block diagonal matrix is easily obtained when the block sizes are
small.

\section{Conditional simulations}\label{condsimsection}

Here we describe how we conditionally simulate the process at the
unobserved sites, $\bmu_{01}$ and $\bmu_{02}$. To account for some of
the uncertainty in the fitted model, we construct $\tilde{A}$ with
weights $\tilde{w}_{kb}$ that are sampled from the asymptotic
distribution of the maximum likelihood weights, $\hat{w}_{kb}$, whose
covariance is given by the inverse Hessian of the loglikelihood. We
simulate the partially differenced process in the spectral domain by
jointly simulating the complex Fourier coefficients $\mathbf{Z}_{01}$ and
$\mathbf{Z}_{02}$ from the multivariate complex normal distribution
conditional on the coefficients from the other sites, $(\widehat{\mathbf
{Z}}_{1},\ldots,\widehat{\mathbf{Z}}_{n})$, where the coefficients at
the same frequency from different sites are related with the spatial
coherence function, and coefficients at differing frequencies are
uncorrelated. The partially\vspace*{2pt} differenced process is then constructed
with the transformation $\Delta_{\hat{\alpha}} \mathbf{Y}_{0j} =
C_T(\tilde{A})_{0j}\mathbf{Z}_{0j}$, with $j=1,2$.

\begin{figure}

\includegraphics{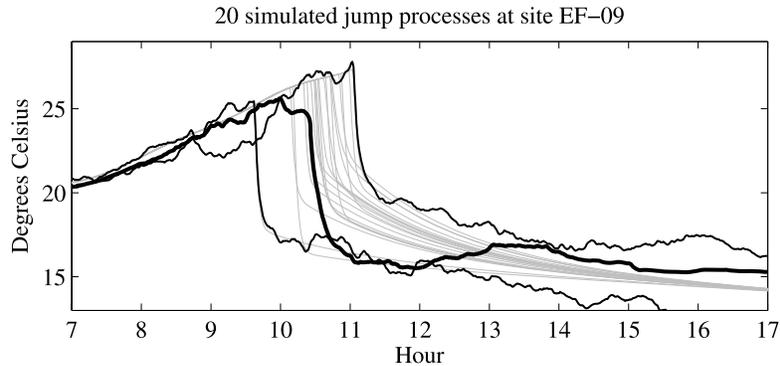}

\caption{20 conditionally simulated jump processes (gray), along with
the data (thick black line) and 2 full conditional simulations (thin
black lines) corresponding to the earliest and latest simulated jump
times among the 20.}
\label{simjump}
\end{figure}

As part of fitting the jump process to the observed sites, we estimated
three parameters for each site: $\tau(\bmu)$, the time of the jump at
site $\bmu$; $D(\bmu)$, the size of the jump; and $\lambda(\bmu)$,
the steepness of the jump. To interpolate the jump process at the
unobserved sites, we model the three parameters, $\tau(\bmu)$, $\log
D(\bmu)$ and $\log1/\lambda(\bmu)$, as independent intrinsic
stationary spatial Gaussian processes with mean functions $d_k - \theta
_k \bmu^\prime\bolds{\phi}_k$ (which can be written as a linear
function of the coordinates) and generalized covariance functions
$G_k(d) = -\eta_k d$ [$k = 1,2,3$ refers to the parameters $\tau(\bmu
)$, $D(\bmu)$, $\lambda(\bmu)$]. The best linear unbiased predictors
of the jump process parameters depend only on contrasts of the observed
jump process parameters, so we may use restricted maximum likelihood to
estimate $\eta_k$ and to predict the parameters at the unobserved
sites [see, e.g., \citet{steinbook}, page 171]. We conditionally
simulate the jump process parameters at the unobserved sites using
bivariate $t$ distributions on 10 degrees of freedom (13 observed
sites---3 mean parameters) to account for uncertainty in the estimates of
$\eta_k$. We then reconstruct the jump process from the conditionally
simulated jump process parameters. As an example, we plot in Figure
\ref{simjump} the data from October 5 at site EF-09, along with 20
conditional simulations of the jump process and 2 conditional
simulations of the temperature process.

We also model the spatial mean function as a spatial Gaussian process
(independent of the other processes). Its generalized covariance has
the same form as that of the jump process parameters, and its mean
function is linear in latitude and elevation. We use the fitted model
to simulate $(s(\bmu_{01}),s(\bmu_{02}))$ jointly and conditionally
on the means from the other sites. The conditionally simulated process
at site $\bmu_{0j}$ is then constructed with
\begin{eqnarray*}
\widehat{Y}(1,\bmu_{0j}) &=& 0,
\\
\widehat{Y}(t,\bmu_{0j}) &=& \hat{\alpha}\widehat{Y}(t-1,\bmu
_{0j}) + \Delta_{\hat{\alpha}}\widehat{Y}(t,\bmu_{0j}),\qquad t>1,
\\
\widehat{X}(t,\bmu_{0j}) &=& \widehat{Y}(t,\bmu_{0j}) +
\hat{m}(t) + s(\bmu_{0j}) + \widehat{J}(t,\bmu_{0j}).
\end{eqnarray*}
We simulate 99 conditionally independent bivariate time series using
the methods described above.\vadjust{\goodbreak} The simulations are very fast; each
bivariate simulation takes just a few seconds on a single processor.
Confidence bands may be constructed by computing the quantiles at each
time point among the conditional simulations. To evaluate our
conditional simulations, we constructed 90\% confidence bands and found
that they had a coverage rate of 89.6\% for the central site and a
coverage rate of 93.3\% for the peripheral site. The roughly accurate
coverage rates are a promising result, although we are slightly
conservative for the peripheral site. Table \ref{conftable} shows that
the predicted temperatures at the peripheral site and during the
%
\begin{table}[b]
\tablewidth=221pt
\caption{Average width of 90\% confidence intervals in $^\circ\mbox
{C}$ for
predicted temperatures at the two unobserved sites over daytime blocks,
nighttime blocks and overall. Sample standard deviations are given in
parentheses}\label{conftable}
\begin{tabular*}{\tablewidth}{@{\extracolsep{\fill}}lccc@{}}
\hline
\multicolumn{1}{c}{}&\multicolumn{1}{c}{\textbf{Daytime}}&\multicolumn
{1}{c}{\textbf{Nighttime}}&\multicolumn{1}{c@{}}{\textbf{Overall}}\\
\hline
Peripheral& 5.95 (1.23) & 4.79 (1.08) & 5.14 (1.24)\\
Central & 5.03 (1.17) & 3.90 (1.00) & 4.24 (1.18)\\
\hline
\end{tabular*}
\end{table}
%
%
\begin{figure}

\includegraphics{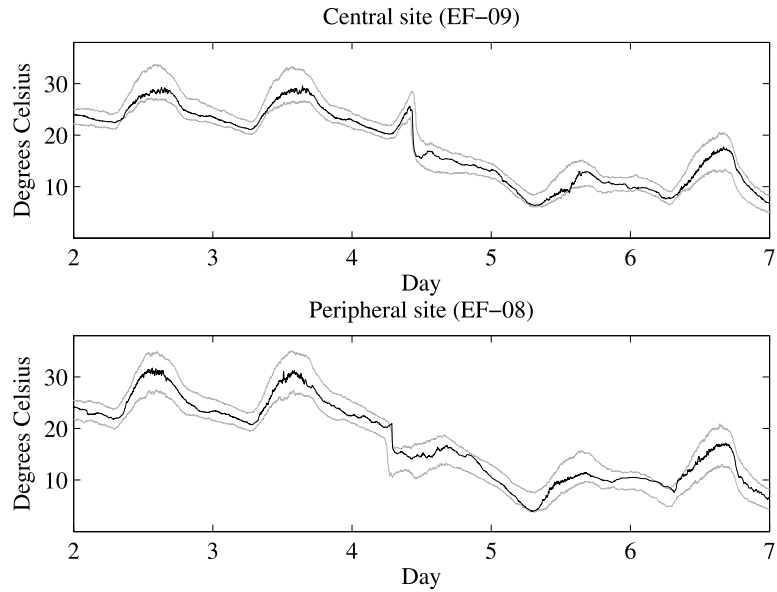}

\caption{90\% confidence bands for predicted temperature at the two
held-out sites over days 3 through 7. This time period includes the
extreme drop in temperature, which occurred on day 5.}
\label{conf01}
\end{figure}
%
\begin{figure}

\includegraphics{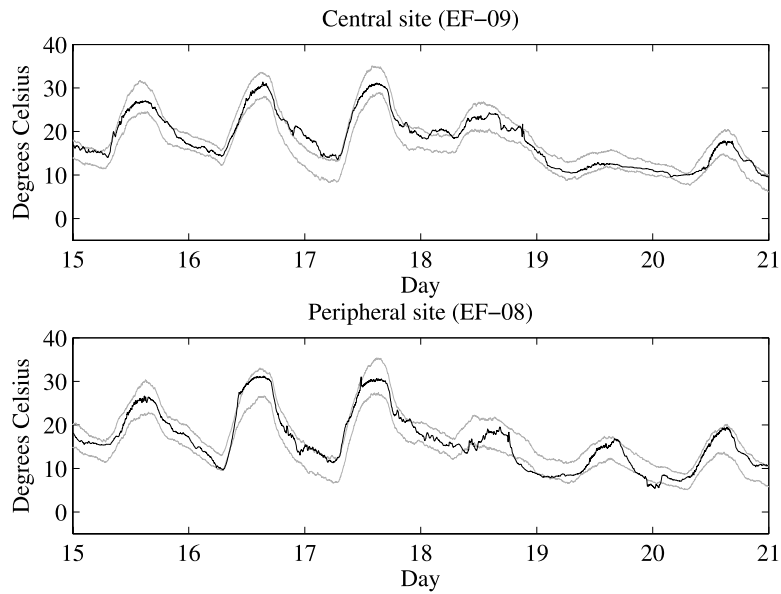}

\caption{90\% confidence bands for predicted temperature at the two
held-out sites over days 16 through 21. The confidence bands are
usually wider during the daytime, and the peripheral site's confidence
band is generally wider than that of the central site. }
\label{conf02}
\end{figure}
daytime have wider confidence bands. This is not surprising, as we
expect it to be more difficult to predict temperatures at the boundary
of the observation region, and the daytime temperatures tend to be more
variable. The confidence bands around the time of the jump become very
wide; at the peripheral site, which undergoes a smaller jump because
the drop occurs earlier in the day, the confidence band is wider than
8$^\circ\mbox{C}$ for more than 40 minutes, and at the central site, the
confidence band is wider than 9.5$^\circ\mbox{C}$ for 30 minutes near
the time of
its drop. In Figures \ref{conf01} and \ref{conf02} we plot a portion
of the data from the unobserved sites along with the 90\% confidence
bands for predicted temperature.

\section{Conclusions}

We have proposed a spatial--temporal model that aims to capture
nonstationary variance and correlation in a set of high frequency
temperature data. The model also allows for the variance to depend on
another meteorological covariate, solar radiation, and includes
spatial--temporal jumps. We provided computationally efficient methods
for fitting the model to a large data set and for generating
spatial--temporal simulations from the model, conditional on the
observations. Most of the computational effort is spent on fitting the
model. Once the model is fit, the conditional simulations can be
computed very quickly. The conditional simulations result in a suite of
temperature values at unobserved locations, and we have shown that the
simulated data reflect some of the uncertainties that we expect,
namely, that the interpolations are more uncertain during the daytime,
that there is more uncertainty in predicting at a peripheral location
than there is in predicting at an interior location, and the
uncertainties are inflated near the time of the jump.

\begin{appendix}\label{app}

\section{Parametric form and fitting of jump process}\label{appA}

As seen in Figure \ref{day5undiff}, the temperatures at each site
follow the usual diurnal cycle until the jump occurs, when they undergo
a very sharp drop followed by a slow decay toward a fixed temperature,
roughly 15$^\circ\mbox{C}$. This suggests that the following may be an
appropriate formulation of the jump process:
%
\begin{equation}
J(t,\bmu) = \cases{ b_1(t) - m(t), &\quad if $4 \times1440 < t < \tau({
\bmu})$,
\cr
b^*(t,\bmu), &\quad if $\tau({\bmu}) \leq t \leq5 \times1440$,
\cr
0, &\quad
otherwise,}
\end{equation}
where $b_1$ is a mean function that shadows the diurnal cycle, and
%
\begin{eqnarray}
b^*(t,\bmu) &=& \bigl[b_1\bigl(\tau({\bmu})\bigr) - m\bigl(\tau({
\bmu})\bigr)\bigr]\exp\bigl(-\nu_1\bigl(t-\tau({\bmu})\bigr)\bigr)
\nonumber
\\
&&{}- D({\bmu})\int_0^{t-\tau({\bmu})} \frac{\lambda({\bmu
})^{\beta}}{\Gamma(\beta)}x^{\beta-1}
\exp\bigl(-\lambda({\bmu})x\bigr) \,dx
\nonumber\\[-8pt]\\[-8pt]
&&{}+ D({\bmu})\bigl[1 - \exp\bigl(-\nu_2\bigl(t-\tau({\bmu})\bigr)
\bigr)\bigr]
\nonumber
\\
&&{}+ \bigl[b_2(t) - m(t)\bigr] \bigl[1-\exp\bigl(-\nu_3
\bigl(t-\tau({\bmu})\bigr)\bigr)\bigr].
\nonumber
\end{eqnarray}
The integral term is simply the incomplete gamma function with rate
$\lambda({\bmu})$ and shape $\beta$. We use the incomplete gamma
function to model the drop because the first differences of temperature
immediately following the jump resemble gamma densities. Additionally,
we could allow the shape parameter, $\beta$, to vary with location,
but for simplicity, we assume that $\beta$ is constant in space and
fix it at $\beta= 1.3$. The function $b_2$ is a post-jump mean that we
take to be linear. The rates $\nu_1,\nu_2$ and $\nu_3$ are less
critical, so they may be taken to be equal, but they should be much
smaller than $\lambda({\bmu})$. In our analysis, we assume $\nu
_j=0.01$ for each $j$.

To fit $b_1(t)$, we compute a moving average of the average temperature
at time~$t$, where the average is taken over the sites that have not
undergone a jump up to time~$t$. Thus, for each $\bmu$, we must find a
preliminary estimate of $\tau({\bmu})$, which we call $\overline
{\tau}({\bmu})$, found by taking the first time at each site for
which the first differences are below $-0.35^\circ\mbox{C}$. Later,
the estimates
of the jump times will be refined. To estimate $b_2(t)$, we fit a
linear function to the average of the temperatures at time $t$ from the
sites for which $\overline{\tau}({\bmu}) + 60 < t$. The line is
constrained so that $b_2(1440\times5+1) = m(1440\times5+1)$, the
value of the temporal mean function at the start of the sixth day. To
fit $\tau(\bmu)$, $D(\bmu)$ and $\lambda({\bmu})$, we minimize the
sum of squared differences between the first differences of the fitted
jump process with those parameters and the first differences of
temperature, where the sum of squares is taken over all sites and for
times $\overline{\tau}({\bmu})-3,\ldots,\overline{\tau}({\bmu})+20$.

\section{Choosing the regimes}\label{appB}

One problem with the average daytime and average nighttime periodograms
is that the periodograms are taken over blocks of data that are never
longer than 18 hours, so they do not give any information about the
very low frequency (less than 1 or 2 cycles per day) behavior of the
process. Therefore, we are forced to try to learn something about the
low frequency behavior from periodograms taken over longer blocks of
data. We estimate the shape of $A$ at low frequencies $(-\omega
_S,\omega_S)$ by computing the periodogram using the entire month of
data at each site and averaging the periodograms over the sites, which
we call the average monthly periodogram. Here, we take $\omega_S = 2$
cycles per day.

Based on the nonmonotonic shape of the daytime and nighttime
periodograms and the requirement that the regimes be positive, it seems
reasonable to fit cubic B-splines to one half of the log of the
periodograms to model the regimes. The knots are chosen manually, and
we take them to be more dense at lower frequencies. The positions of
the knots are $(1/3, 2/3, 1, 4/3, 5/3,\break 2, 4, 8, 12, 24, 60, 120, 360)$ in
units of cycles per day. Under the constraints that the first
derivative of $\mu_k(\omega)$ at $\omega= 0$ and $\pi$ be equal to
zero, and $\mu_k(\omega) = \mu_k(-\omega)$, we are left with 15
basis functions per regime in the cubic B-spline representation, of
which 8 are nonzero in the interval $(-\omega_S,\omega_S)$. We fit
the coefficients for those 8 basis functions by minimizing the sum of
squares between the associated B-spline and one half the log of the
average monthly periodogram at the frequencies in the interval
$(-\omega_S,\omega_S)$. Holding those 8 coefficients constant at
their least squares estimates, we choose the remaining 7 coefficients
for each regime by minimizing the sum of squares between the associated
B-spline and one half the log of the average daytime (and nighttime)
periodogram. This procedure ensures that $\mu_1$ and $\mu_2$ are
equal in the interval $(-\omega_S,\omega_S)$ but free to vary at
higher frequencies. 

\section{Motivation for partial differencing}\label{appC}

Until now, we have given only a brief explanation for the need for
partial differencing. If we set $\alpha=1$ and perform the analysis on
the first differences of temperature, the conditional simulations
become almost useless. As the month progresses, the simulated
temperatures tend to drift away from the data, resulting in very wide
confidence bands for most of the month. This is a problem that was not
observed in the analysis in \citet{stein09}, in which he fit and
simulated from a smooth, uniformly modulated model. Here, the
undifferencing operation interacts with the discontinuous jumps in $A$,
causing the undifferenced simulations to drift. More specifically, in
the stationary case the basis functions for the process are
trigonometric functions, which integrate to zero over every cycle. In
the locally stationary case, the basis functions are
amplitude-modulated trigonometric functions. When $A$ is smooth as a
function of time, as in \citet{stein09}, all but the highest frequency
functions may approximately integrate to zero, but when $A$ has jumps,
this is no longer the case, so when the simulations are undifferenced,
the results can be quite unpredictable. Partial differencing appears to
solve the problem and has some theoretical merit given the nature of
the temperature process, as discussed in Section~\ref{lstatsection}.
\end{appendix}



%

\printaddresses


\begin{thebibliography}{29}

\bibitem[\protect\citeauthoryear{Anderes and Chatterjee}{2009}]{anderes09}
%
\begin{barticle}[mr]
\bauthor{\bsnm{Anderes},~\bfnm{Ethan}\binits{E.}} \AND
\bauthor{\bsnm{Chatterjee},~\bfnm{Sourav}\binits{S.}}
(\byear{2009}).
\btitle{Consistent estimates of deformed isotropic {G}aussian random
fields on
the plane}.
\bjournal{Ann. Statist.}
\bvolume{37}
\bpages{2324--2350}.
\bid{doi={10.1214/08-AOS647}, issn={0090-5364}, mr={2543694}}
\bptok{imsref}%
\end{barticle}
%
\endbibitem

\bibitem[\protect\citeauthoryear{Anderes and Stein}{2008}]{anderes08}
%
\begin{barticle}[mr]
\bauthor{\bsnm{Anderes},~\bfnm{Ethan~B.}\binits{E.~B.}} \AND
\bauthor{\bsnm{Stein},~\bfnm{Michael~L.}\binits{M.~L.}}
(\byear{2008}).
\btitle{Estimating deformations of isotropic {G}aussian random fields
on the
plane}.
\bjournal{Ann. Statist.}
\bvolume{36}
\bpages{719--741}.
\bid{doi={10.1214/009053607000000893}, issn={0090-5364}, mr={2396813}}
\bptok{imsref}%
\end{barticle}
%
\endbibitem

\bibitem[\protect\citeauthoryear{Arya}{2001}]{aryabook}
%
\begin{bbook}[author]
\bauthor{\bsnm{Arya},~\bfnm{S.~Pal}\binits{S.~P.}}
(\byear{2001}).
\btitle{Introduction to Micrometeorology},
\bedition{2nd} ed.
\bpublisher{Academic Press}, \blocation{San Diego, CA}.
\bptok{imsref}%
\end{bbook}
%
\endbibitem

\bibitem[\protect\citeauthoryear{Benth and
{\v{S}}altyt{\.e}~Benth}{2007}]{benth07}
%
\begin{barticle}[mr]
\bauthor{\bsnm{Benth},~\bfnm{Fred~Espen}\binits{F.~E.}} \AND
\bauthor{\bsnm{{\v{S}}altyt{\.e}~Benth},~\bfnm{J{\=u}rat{\.e}}\binits{J.}}
(\byear{2007}).
\btitle{The volatility of temperature and pricing of weather derivatives}.
\bjournal{Quant. Finance}
\bvolume{7}
\bpages{553--561}.
\bid{doi={10.1080/14697680601155334}, issn={1469-7688}, mr={2358919}}
\bptok{imsref}%
\end{barticle}
%
\endbibitem

\bibitem[\protect\citeauthoryear{Campbell and Diebold}{2005}]{campbell05}
%
\begin{barticle}[mr]
\bauthor{\bsnm{Campbell},~\bfnm{Sean~D.}\binits{S.~D.}} \AND
\bauthor{\bsnm{Diebold},~\bfnm{Francis~X.}\binits{F.~X.}}
(\byear{2005}).
\btitle{Weather forecasting for weather derivatives}.
\bjournal{J. Amer. Statist. Assoc.}
\bvolume{100}
\bpages{6--16}.
\bid{doi={10.1198/016214504000001051}, issn={0162-1459}, mr={2166065}}
\bptok{imsref}%
\end{barticle}
%
\endbibitem

\bibitem[\protect\citeauthoryear{Dahlhaus}{1996}]{dahlhaus96}
%
\begin{barticle}[mr]
\bauthor{\bsnm{Dahlhaus},~\bfnm{R.}\binits{R.}}
(\byear{1996}).
\btitle{On the {K}ullback--{L}eibler information divergence of locally
stationary processes}.
\bjournal{Stochastic Process. Appl.}
\bvolume{62}
\bpages{139--168}.
\bid{doi={10.1016/0304-4149(95)00090-9}, issn={0304-4149}, mr={1388767}}
\bptok{imsref}%
\end{barticle}
%
\endbibitem

\bibitem[\protect\citeauthoryear{Dahlhaus}{1997}]{dahlhaus97}
%
\begin{barticle}[mr]
\bauthor{\bsnm{Dahlhaus},~\bfnm{R.}\binits{R.}}
(\byear{1997}).
\btitle{Fitting time series models to nonstationary processes}.
\bjournal{Ann. Statist.}
\bvolume{25}
\bpages{1--37}.
\bid{doi={10.1214/aos/1034276620}, issn={0090-5364}, mr={1429916}}
\bptok{imsref}%
\end{barticle}
%
\endbibitem

\bibitem[\protect\citeauthoryear{Dahlhaus}{2000}]{dahlhaus00}
%
\begin{barticle}[mr]
\bauthor{\bsnm{Dahlhaus},~\bfnm{Rainer}\binits{R.}}
(\byear{2000}).
\btitle{A likelihood approximation for locally stationary processes}.
\bjournal{Ann. Statist.}
\bvolume{28}
\bpages{1762--1794}.
\bid{doi={10.1214/aos/1015957480}, issn={0090-5364}, mr={1835040}}
\bptok{imsref}%
\end{barticle}
%
\endbibitem

\bibitem[\protect\citeauthoryear{Dahlhaus}{2012}]{dahlhaus12}
%
\begin{bincollection}[author]
\bauthor{\bsnm{Dahlhaus},~\bfnm{Rainer}\binits{R.}}
(\byear{2012}).
\btitle{Locally stationary processes}.
In \bbooktitle{Handbook of Statistics 30, Time Series Analysis: Methods
and Applications}
\bvolume{30}
\bpages{351--408}.
\bpublisher{Elsevier}, \blocation{Amsterdam}.
\bptok{imsref}%
\end{bincollection}
%
\endbibitem

\bibitem[\protect\citeauthoryear{Daly, Neilson and Phillips}{1994}]{daly94}
%
\begin{barticle}[author]
\bauthor{\bsnm{Daly},~\bfnm{Christopher}\binits{C.}},
\bauthor{\bsnm{Neilson},~\bfnm{Ronald~P.}\binits{R.~P.}} \AND
\bauthor{\bsnm{Phillips},~\bfnm{Donald~L.}\binits{D.~L.}}
(\byear{1994}).
\btitle{A statistical-topographic model for mapping climatological
precipitation over mountainous terrain}.
\bjournal{Journal of Applied Meteorology}
\bvolume{33}
\bpages{140--158}.
\bptok{imsref}%
\end{barticle}
%
\endbibitem

\bibitem[\protect\citeauthoryear{Fuentes}{2002}]{fuentes02}
%
\begin{barticle}[mr]
\bauthor{\bsnm{Fuentes},~\bfnm{Montserrat}\binits{M.}}
(\byear{2002}).
\btitle{Spectral methods for nonstationary spatial processes}.
\bjournal{Biometrika}
\bvolume{89}
\bpages{197--210}.
\bid{doi={10.1093/biomet/89.1.197}, issn={0006-3444}, mr={1888368}}
\bptok{imsref}%
\end{barticle}
%
\endbibitem

\bibitem[\protect\citeauthoryear{Fuentes, Chen and Davis}{2008}]{fuentes07}
%
\begin{barticle}[mr]
\bauthor{\bsnm{Fuentes},~\bfnm{Montserrat}\binits{M.}},
\bauthor{\bsnm{Chen},~\bfnm{Li}\binits{L.}} \AND
\bauthor{\bsnm{Davis},~\bfnm{Jerry~M.}\binits{J.~M.}}
(\byear{2008}).
\btitle{A class of nonseparable and nonstationary spatial temporal covariance
functions}.
\bjournal{Environmetrics}
\bvolume{19}
\bpages{487--507}.
\bid{doi={10.1002/env.891}, issn={1180-4009}, mr={2523910}}
\bptnote{check year}%
\bptok{imsref}%
\end{barticle}
%
\endbibitem

\bibitem[\protect\citeauthoryear{Gardner, Napolitano and
Paura}{2006}]{gardner06}
%
\begin{barticle}[author]
\bauthor{\bsnm{Gardner},~\bfnm{William~A.}\binits{W.~A.}},
\bauthor{\bsnm{Napolitano},~\bfnm{Antonio}\binits{A.}} \AND
\bauthor{\bsnm{Paura},~\bfnm{Luigi}\binits{L.}}
(\byear{2006}).
\btitle{Cyclostationarity: Half a century of research}.
\bjournal{Signal Processing}
\bvolume{86}
\bpages{639--697}.
\bptok{imsref}%
\end{barticle}
%
\endbibitem

\bibitem[\protect\citeauthoryear{Guinness and Stein}{2013}]{guinness12}
%
\begin{barticle}[author]
\bauthor{\bsnm{Guinness},~\bfnm{Joseph}\binits{J.}} \AND
\bauthor{\bsnm{Stein},~\bfnm{Michael~L.}\binits{M.~L.}}
(\byear{2013}).
\btitle{Transformation to approximate independence for locally stationary
Gaussian processes}.
\bjournal{J. Time Series Anal.}
\bvolume{34}
\bpages{574--590}.
\bptok{imsref}%
\end{barticle}
%
\endbibitem

\bibitem[\protect\citeauthoryear{Haslett and Raftery}{1989}]{haslett89}
%
\begin{barticle}[author]
\bauthor{\bsnm{Haslett},~\bfnm{John}\binits{J.}} \AND
\bauthor{\bsnm{Raftery},~\bfnm{Adrian~E.}\binits{A.~E.}}
(\byear{1989}).
\btitle{Space--time modelling with long-memory dependence: Assessing Ireland's
wind power resource}.
\bjournal{J. Appl. Stat.}
\bvolume{38}
\bpages{1--50}.
\bptok{imsref}%
\end{barticle}
%
\endbibitem

\bibitem[\protect\citeauthoryear{Higdon, Swall and Kern}{1999}]{higdon99}
%
\begin{bmisc}[author]
\bauthor{\bsnm{Higdon},~\bfnm{David}\binits{D.}},
\bauthor{\bsnm{Swall},~\bfnm{Jenise}\binits{J.}} \AND
\bauthor{\bsnm{Kern},~\bfnm{John}\binits{J.}}
(\byear{1999}).
\bhowpublished{Non-stationary spatial modeling.
In \textit{Bayesian Statistics} 6: \textit{Proceedings of the Sixth Valencia
International Meeting}
761--768.
Oxford Univ. Press, New York}.
\bptok{imsref}%
\end{bmisc}
%
\endbibitem

\bibitem[\protect\citeauthoryear{Hijmans et~al.}{2005}]{hijmans05}
%
\begin{barticle}[author]
\bauthor{\bsnm{Hijmans},~\bfnm{Robert~J.}\binits{R.~J.}},
\bauthor{\bsnm{Cameron},~\bfnm{Susain~E.}\binits{S.~E.}},
\bauthor{\bsnm{Parra},~\bfnm{Juan~L.}\binits{J.~L.}},
\bauthor{\bsnm{Jones},~\bfnm{Peter~G.}\binits{P.~G.}} \AND
\bauthor{\bsnm{Jarvis},~\bfnm{Andy}\binits{A.}}
(\byear{2005}).
\btitle{Very high resolution interpolated climate surfaces for global land
areas}.
\bjournal{International Journal of Climatology}
\bvolume{25}
\bpages{1965--1978}.
\bptok{imsref}%
\end{barticle}
%
\endbibitem

\bibitem[\protect\citeauthoryear{Lantu{\'{e}}joul}{2002}]{geosimbook}
%
\begin{bbook}[author]
\bauthor{\bsnm{Lantu{\'{e}}joul},~\bfnm{Christian}\binits{C.}}
(\byear{2002}).
\btitle{Geostatistical Simulation: Models and Algorithms}.
\bpublisher{Springer}, \blocation{New York}.
\bptok{imsref}%
\end{bbook}
%
\endbibitem

\bibitem[\protect\citeauthoryear{Paciorek and Schervish}{2006}]{paciorek06}
%
\begin{barticle}[mr]
\bauthor{\bsnm{Paciorek},~\bfnm{Christopher~J.}\binits{C.~J.}} \AND
\bauthor{\bsnm{Schervish},~\bfnm{Mark~J.}\binits{M.~J.}}
(\byear{2006}).
\btitle{Spatial modelling using a new class of nonstationary covariance
functions}.
\bjournal{Environmetrics}
\bvolume{17}
\bpages{483--506}.
\bid{doi={10.1002/env.785}, issn={1180-4009}, mr={2240939}}
\bptok{imsref}%
\end{barticle}
%
\endbibitem

\bibitem[\protect\citeauthoryear{Priestley}{1965}]{priestley65}
%
\begin{barticle}[mr]
\bauthor{\bsnm{Priestley},~\bfnm{M.~B.}\binits{M.~B.}}
(\byear{1965}).
\btitle{Evolutionary spectra and non-stationary processes (with discussion)}.
\bjournal{J.~R. Stat. Soc. Ser. B Stat. Methodol.}
\bvolume{27}
\bpages{204--237}.
\bid{issn={0035-9246}, mr={0199886}}
\bptnote{check related}%
\bptok{imsref}%
\end{barticle}
%
\endbibitem

\bibitem[\protect\citeauthoryear{Priestley}{1981}]{priestleybook}
%
\begin{bbook}[author]
\bauthor{\bsnm{Priestley},~\bfnm{Maurice~B.}\binits{M.~B.}}
(\byear{1981}).
\btitle{Spectral Analysis and Time Series}.
\bpublisher{Academic Press}, \blocation{San Diego}.
\bnote{Seventh printing, 1992}.
\bptok{imsref}%
\end{bbook}
%
\endbibitem

\bibitem[\protect\citeauthoryear{Rubin}{1987}]{rubinbook}
%
\begin{bbook}[mr]
\bauthor{\bsnm{Rubin},~\bfnm{Donald~B.}\binits{D.~B.}}
(\byear{1987}).
\btitle{Multiple Imputation for Nonresponse in Surveys}.
\bpublisher{Wiley}, \blocation{New York}.
\bid{doi={10.1002/9780470316696}, mr={0899519}}
\bptok{imsref}%
\end{bbook}
%
\endbibitem

\bibitem[\protect\citeauthoryear{Sampson and Guttorp}{1992}]{sampson92}
%
\begin{barticle}[author]
\bauthor{\bsnm{Sampson},~\bfnm{Paul~D.}\binits{P.~D.}} \AND
\bauthor{\bsnm{Guttorp},~\bfnm{Peter}\binits{P.}}
(\byear{1992}).
\btitle{Nonparametric estimation of nonstationary spatial covariance
structure}.
\bjournal{J. Amer. Statist. Assoc.}
\bvolume{87}
\bpages{108--119}.
\bptok{imsref}%
\end{barticle}
%
\endbibitem

\bibitem[\protect\citeauthoryear{Schneider}{2006}]{schneider06}
%
\begin{barticle}[author]
\bauthor{\bsnm{Schneider},~\bfnm{Tapio}\binits{T.}}
(\byear{2006}).
\btitle{Analysis of incomplete data: Readings from the statistics literature}.
\bjournal{Bulletin of the American Meteorological Society}
\bvolume{87}
\bpages{1410--1411}.
\bptok{imsref}%
\end{barticle}
%
\endbibitem

\bibitem[\protect\citeauthoryear{Stein}{1999}]{steinbook}
%
\begin{bbook}[mr]
\bauthor{\bsnm{Stein},~\bfnm{Michael~L.}\binits{M.~L.}}
(\byear{1999}).
\btitle{Interpolation of Spatial Data: Some Theory for Kriging}.
\bpublisher{Springer}, \blocation{New York}.
\bid{doi={10.1007/978-1-4612-1494-6}, mr={1697409}}
\bptok{imsref}%
\end{bbook}
%
\endbibitem

\bibitem[\protect\citeauthoryear{Stein}{2005}]{stein05}
%
\begin{barticle}[mr]
\bauthor{\bsnm{Stein},~\bfnm{Michael~L.}\binits{M.~L.}}
(\byear{2005}).
\btitle{Statistical methods for regular monitoring data}.
\bjournal{J. R. Stat. Soc. Ser. B Stat. Methodol.}
\bvolume{67}
\bpages{667--687}.
\bid{doi={10.1111/j.1467-9868.2005.00520.x}, issn={1369-7412}, mr={2210686}}
\bptok{imsref}%
\end{barticle}
%
\endbibitem

\bibitem[\protect\citeauthoryear{Stein}{2009}]{stein09}
%
\begin{barticle}[mr]
\bauthor{\bsnm{Stein},~\bfnm{Michael~L.}\binits{M.~L.}}
(\byear{2009}).
\btitle{Spatial interpolation of high-frequency monitoring data}.
\bjournal{Ann. Appl. Stat.}
\bvolume{3}
\bpages{272--291}.
\bid{doi={10.1214/08-AOAS208}, issn={1932-6157}, mr={2668708}}
\bptok{imsref}%
\end{barticle}
%
\endbibitem

\bibitem[\protect\citeauthoryear{Whittle}{1953}]{whittle53}
%
\begin{barticle}[mr]
\bauthor{\bsnm{Whittle},~\bfnm{P.}\binits{P.}}
(\byear{1953}).
\btitle{Estimation and information in stationary time series}.
\bjournal{Ark. Mat.}
\bvolume{2}
\bpages{423--434}.
\bid{issn={0004-2080}, mr={0060797}}
\bptok{imsref}%
\end{barticle}
%
\endbibitem

\end{thebibliography}
\end{document}